\documentclass{sig-alternate}

\usepackage{url}
\usepackage{xspace}

\newcommand{\eg}{\textit{e.g.,}\xspace}

\renewcommand{\paragraph}[1]{\vspace{.2cm} \noindent \textbf{#1}}

\begin{document}

\title{Enabling Reproducible Science with VisTrails}

\numberofauthors{3} %
\author{
\alignauthor
David Koop\\
       \affaddr{Polytechnic Institute of}\\
       \affaddr{New York University}\\
       \affaddr{New York, NY, USA}\\
       \email{dakoop@nyu.edu}
\alignauthor
Juliana Freire\\
       \affaddr{Polytechnic Institute of}\\
       \affaddr{New York University}\\
       \affaddr{New York, NY, USA}\\
       \email{juliana.freire@nyu.edu}
\alignauthor Cl\'audio T. Silva\\
       \affaddr{Polytechnic Institute of}\\
       \affaddr{New York University}\\
       \affaddr{New York, NY, USA}\\
       \email{csilva@nyu.edu}
}
\date{17 November 2013}

\maketitle
\begin{abstract}
  With the increasing amount of data and use of computation in
  science, software has become an important component in many
  different domains.  Computing is now being used more often and in
  more aspects of scientific work including data acquisition,
  simulation, analysis, and visualization.  To ensure reproducibility,
  it is important to capture the different computational processes
  used as well as their executions.  VisTrails is an open-source
  scientific workflow system for data analysis and visualization that
  seeks to address the problem of integrating varied tools as well as
  automatically documenting the methods and parameters employed.
  Growing from a specific project need to supporting a wide array of
  users required close collaborations in addition to new research
  ideas to design a usable and efficient system.  The VisTrails
  project now includes standard software processes like unit testing
  and developer documentation while serving as a base for further research.
  In this paper, we describe how VisTrails has developed and how our
  efforts in structuring and advertising the system have contributed
  to its adoption in many domains.
\end{abstract}

\section{Introduction}
\label{sec:intro}

As scientists adopt computational methods for their work, their
need for software tools has been growing.  Not only is the amount of
data available exploding, but techniques like computational simulation
and modeling have become more prevalent in many domains.  These
changes have made it, in many cases, much easier to generate, test,
and validate new hypotheses, and they have also introduced new issues
and needs.  Not only has this evolution in science created the need
for new computer science research, software, and infrastructure, but
it has also created new opportunities for collaboration and outreach
between computing and other scientific domains.  The constant pursuit
of improved hardware, more storage, and faster algorithms surely
contributes to furthering computational science, but areas like data
management, provenance, and usability have seen increased visibility
as computation expands its impact.

Cheaper hardware for computation and storage, as well as improved software
have been used to obtain results more
quickly, often allowing analyses that were previously unthinkable.
These analyses are often complex and require many trial-and-error
steps.  Thus, it is important to consider and record each step in the
analysis, archive raw data, and enable other scientists to
independently verify results.  In the drive toward results, computing
has enabled the race to results without always tracking the data and
methods involved.  Compounding this problem, many analyses require the
use of multiple software tools to achieve a result.  For example, raw sensor data
might be cleaned by a pre-processing step before it is run through a
set of scripts that extract the specific quantities of interest and
finally aggregated and plotted.

A hallmark of scientific reasoning and experimentation has been the
exploration of different hypotheses in order to better understand
phenomena.  With more computational steps and a wider array of rapidly
evolving software, the ability to understand and control computational
research often relies on understanding different, often-complex
interfaces.  In some cases, this complexity has led to more rigid
workflows where exploration is more limited.  Finally, while scripts
and other tools can help integrate such workflows, the responsibility
for tracking changes and linking results to the steps used is often
left to the scientist.  Both education and improvements in software
can help address these issues, and they often require collaboration
between scientists and software architects.

In this paper, we discuss how VisTrails, an open-source scientific
workflow system for data analysis and visualization, was created to
enable scientists to integrate existing libraries and new tools while
making exploration and reproducibility easier.  Furthermore, we
discuss how a variety of collaborations with domain scientists have
shaped its development and specific research ideas as well as enabled
scientific results.  With almost a decade of development and use, the
software continues to evolve, and we detail important tasks outside of
programming like documentation and code organization that must be
considered in such a project.  We also discuss how outreach and
education have helped the project grow.  We conclude with a discussion
of some lessons learned and some thoughts about the future.

\section{Genesis \& Background}
\label{sec:genesis}

The vision and development of VisTrails began in 2004, and during the
past decade, there have been significant changes both in direction and
technology.  The initial motivation for VisTrails came from the Center
for Coastal Margin Observation \& Prediction
(CMOP)\footnote{\url{http://www.stccmop.org}} which brings together
oceanographers, environmental scientists, biologists, and chemists to
achieve a better understanding of physical, chemical, and biological
processes regulating river-to-ocean ecosystems. CMOP has amassed vast
amounts of data about the Columbia River, both collected from sensors
and derived by simulations, that needed to be analyzed and
visualized. To help CMOP scientists make sense of these data, IT
personnel, modelers and visualization experts created numerous
computational processes (scripts). The creation of a new visualization
was often a laborious and time consuming task: simulation workflows
were designed and run by the modelers; simulation results were then
given to the visualization expert who assembled the pipelines to
generate visualizations.  This problem was compounded due to the fact
that calibrating the simulation model requires a trial-and-error
process where the scripts need to be combined and iteratively refined.

Significant effort was required to build new workflows and to
manually modify existing workflows to cater to new requirements (\eg to
use different parameters or different visualization techniques).
Besides being challenging for IT personnel and project members with a
computer science background, the complexity involved in this process
made it impossible for domain scientists to manipulate the workflows.
Producing a visualization that could generate insight often required
help from a person familiar with the computation, a visualization
expert, and the scientist(s).  This process often took days and
significant effort from all involved.  Our initial goal with VisTrails
was to assist visualization and computing experts in order to streamline
this process.  However, as the system developed, we realized that by
improving usability and integrating simplified interfaces, we could
also enable scientists to perform some of these tasks themselves.

One of the challenges was integrating a variety of different tools and
libraries in order to generate results.  While shell scripts or
manually recorded protocols could often be used to execute repetitive
analyses, modifying or updating these computations was often
non-trivial.  VisTrails, building on work in workflows and
computational steering, adopted the scientific workflow (pipeline)
representation along with a visual interface for composing and
configuring them.  This allows a uniformity between each computational
module's representation with connections between input and output
ports and settable parameters.  Thus, a new step can be added by
incorporating the new module and appropriately linking it into the
pipeline.  Such functionality enables quicker changes that make
workflows a tool for \emph{exploratory} analysis in addition to
repetitive jobs.

\begin{figure}[t]
\includegraphics[width=0.48\textwidth]{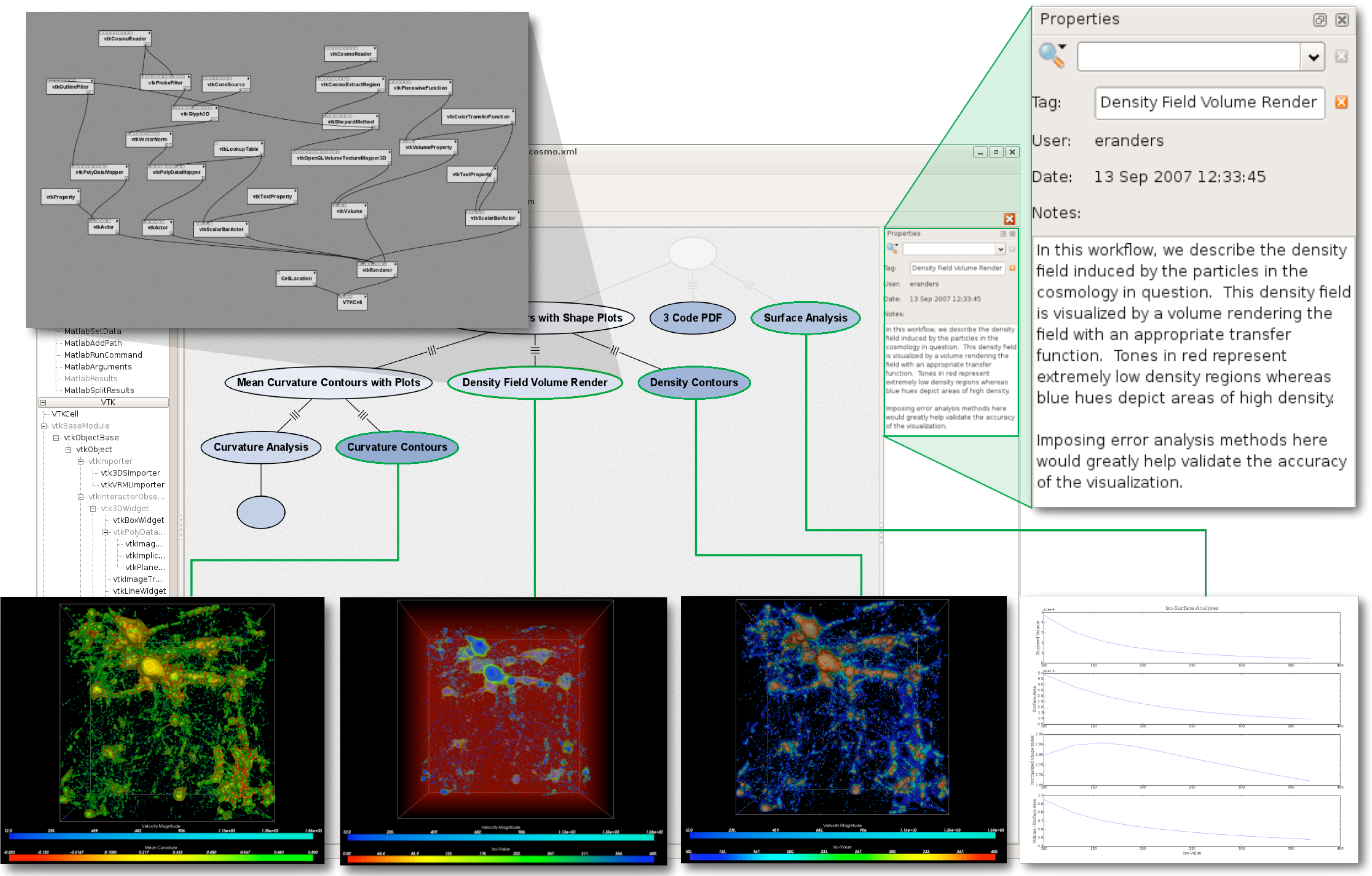}
\caption{VisTrails uses a version tree to show different workflows and
  their relationships.  By selecting a version, users can examine,
  modify, and execute the workflow as well as tag and annotate each version.}
\label{fig:vistrails}
\end{figure}

Another key ingredient in supporting scientific exploration was
keeping track of the changes made during development.  One of the
goals was to ensure that any executed pipeline would be preserved so
that it could be run in the future to reproduce the results.  To make
this unintrusive for users, we decided to automatically capture these
workflows and the refinements applied to them, rather than requiring
users to remember to save (and version) them.  For efficiency, we
introduced \emph{change-based provenance} to compactly capture a
collection of related workflows~\cite{freire:2006:ipaw}.  Because
there were likely few changes (a couple parameter changes, some added
modules) between any two workflows, VisTrails captures the
\emph{actions} that specify how one workflow was transformed into
another.  In contrast to many version control systems, such changes
are defined by the actual actions a user employed to transform the
workflow rather than deduced from the starting and ending workflows
(as a textual diff usually works).  Noting the importance of being
able to quickly return to a previous analysis, VisTrails introduced
the \emph{version tree} to help users navigate all of the past
workflow versions.  This interface enables exploration as users do not
need to worry about saving each version and can easily switch back to
any past waypoint or discarded idea.

\section{Collaborative Impact}
\label{sec:collab}

As the development of VisTrails progressed, we were pleased to attract a
number of users from a diverse set of backgrounds, ranging from
invasive species modeling and climate data analysis to theoretical physics.  Their motivation
for using the system also varied; some users wanted comparative
visualization capabilities while others were most interested in
capturing provenance information.  Not only did these users bring bugs
to our attention and spur improved documentation, but they also helped
shape the development of new features and the improvement of existing
ideas.  As VisTrails has focused on usability concerns, the ability
for users to accomplish their scientific goals has been a focus of our
work.
We have developed many fruitful collaborations that have helped to
sustain VisTrails development which was initially funded by a
relatively small National Science Foundation (NSF) grant.  The
collaborations have both raised the profile of VisTrails and provided
more diverse opportunities for funding.
Below, we highlight some of them.

\paragraph{The Center for Coastal Margin Observation \& Prediction
  (CMOP).} CMOP was one of our original collaborators when the
VisTrails project began.  With an abundance of observed data from the
Columbia River and a wide set of simulations and forecasts, scientists
perform a plethora of analysis and visualization tasks.  Understanding
exactly which data and approaches were used (especially as this data
is continuously gathered over time) is extremely important in
producing accurate conclusions.  VisTrails was used to automate some
of the visualization products using workflows.  As some users found
workflows too complex, we introduced mashups as a Web-based method for
generating visualizations from a user-configurable set of date ranges
and parameters~\cite{santos:2009:tvcg}.

\paragraph{The Algorithms and Libraries for Physics Simulations (ALPS)
  project\footnote{\url{http://alps.comp-phys.org}}.} ALPS is an
open-source effort to provide standard simulation codes for quantum
systems.  The project leaders were interested in providing full
provenance information for simulations and analyses so that results
would be reproducible.  They developed a VisTrails package that
wrapped their C++ libraries as well as a number of examples that
showed how the package could be used.  Because their code evolved
quickly, the need to track different versions of the VisTrails package
was important in adopting the system, and the notion of workflow
upgrades grew from these needs~\cite{koop:2010:ipaw}.  In addition,
the desire to identify and collate input files with the analysis
workflows led to the VisTrails persistence package which not only
provides strong provenance links to data but also incorporates
versioning of input data~\cite{koop:2010:ssdbm}.

\paragraph{Software for Automated Habitat
Modeling (SAHM)\footnote{\url{http://www.fort.usgs.gov/products/software/sahm/}}.}
An invasive species group at the United States Geological Survey
(USGS) used VisTrails to develop SAHM.  With a number of different
models, input layers, regions, and time periods, they are using
VisTrails to track the analyses and
parameters~\cite{morisette:2013:sahm}.  They are also very interested
in comparisons between model outputs when settings are changed,
employing the spreadsheet and a display wall to better explore their
results.  As their work relies on many models written in the R
statistical language, the VisTrails modules call out to the R kernel
to perform some of the calculations.

\begin{figure}[t]
\includegraphics[width=0.48\textwidth]{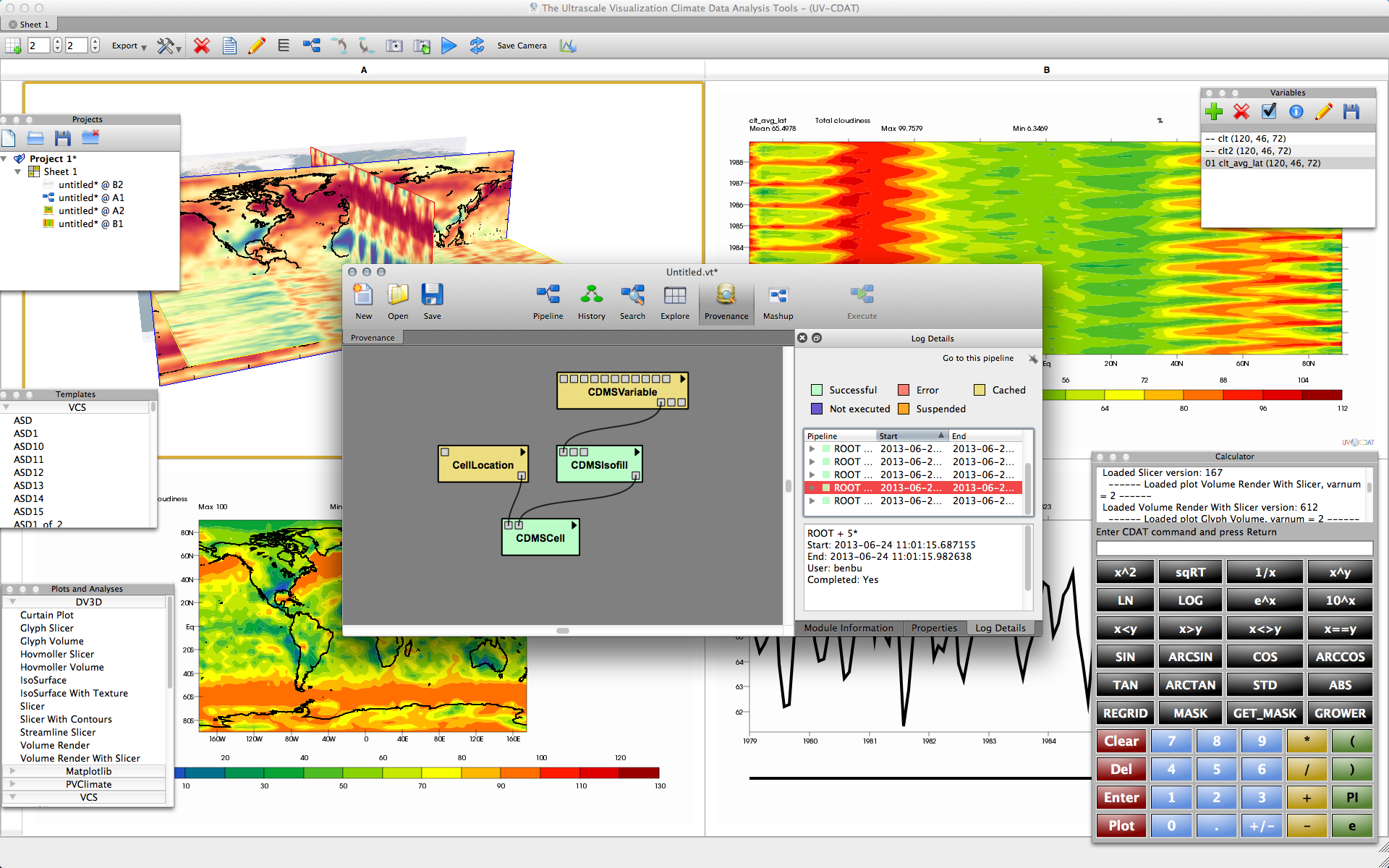}
\vspace{-.3cm}
\caption{The UV-CDAT system uses VisTrails to provide the workflow
  execution and provenance infrastructure with a custom,
  climate-oriented interface.}
\vspace{-.3cm}
\label{fig:uvcdat}
\end{figure}

\paragraph{The Ultra-scale Visualization Climate Data Analysis Tools
  (UV-CDAT) project.} UV-CDAT, led by the Lawrence Livermore National
Laboratory, seeks to deliver a rich set of data exploration and
analysis tools for climate data analysis.  The goal is to make it
easier for climate scientists to locate and analyze data while
capturing provenance information to ensure
reproducibility~\cite{williams:2013:uvcdat}.  They use VisTrails as a
framework and developed a custom graphical user interface (GUI) that
leverages VisTrails workflow execution and provenance capture behind
the scenes~\cite{santos:2012:ipaw}.  For climate scientists, the
distinct operations of loading data, computing derived data, and
generating plots can be better exposed as separate steps in the
interface and transparently composed into workflows.  In addition, the
focus on plots led to the spreadsheet being directly incorporated into
the main window (see Figure~\ref{fig:uvcdat}). Finally, specific
VisTrails packages like
vtDV3D\footnote{\url{http://portal.nccs.nasa.gov/DV3D/}} have been
developed for advanced climate visualization in UV-CDAT.

Each of these collaborations both helped domain scientists in their
work and spurred new ideas and directions for the VisTrails project.
Work with CMOP's many products and parameters generated the idea of
change-based provenance.  The ALPS project's need for integrated data
management and upgradability also led to new VisTrails features.
Because the UV-CDAT project wanted to develop a climate-specific,
visualization-focused interface that also supported workflows and
provenance, we made changes to expose VisTrails as a framework
in addition to a standalone system.  The SAHM group's need for an
interactive workflow led to improved support for custom interface
elements during workflow execution.

\vspace{-.3cm}
\section{Growth \& Development}

VisTrails was initially written in Java, before a C++ version was
released to a few early adopters. It now uses Python, Qt, and PyQt.
With hundreds of thousands of lines of code, we employ many standard
software engineering practices to manage, develop, and test the
system.  In addition, as many of our developers and collaborators are
in different geographic locations, we have worked to better coordinate
work and improve documentation.

\paragraph{Users.}
With over 40,000 downloads since its initial release, VisTrails has
enjoyed significant interest from a variety of communities, and our
web page\footnote{\url{http://www.vistrails.org/}} has seen visitors from over 150 different countries.
While some users have worked in collaboration with us (see
Section~\ref{sec:collab}), others have independently discovered and
used VisTrails.  They come from a variety of disciplines, and have
used it for both business purposes and teaching.  To better support
users, we maintain a user's guide and FAQ in addition to the mailing
lists.  Such support takes significant effort but can also help foster
new collaborations and research ideas.

\paragraph{Code Management.}
Because this project requires collaboration from many geographically dispersed developers,
we use version control to organize code.  After initial prototypes, we
set up a central svn repository for our code but have since moved to
git to gain easier branching and decentralized features.  More
recently, we have utilized github to publish our code.  To track
issues, we have used trac and github, similar to many other projects.  
In addition, we hold weekly developer meetings to discuss important
issues and directions.
Unit tests are included in individual source files with the idea that
each class has a corresponding unit-test class to test its
functionality.  In addition, we have a build machine with multiple
virtual machines that run the test suite to test compatibility with different platforms.

\paragraph{Componentized Architecture}
While one of the original goals of VisTrails was, and still is, to
deliver a fully-functional system that users can install and run with
minimal effort, we also want to make it easier for others to use and
build on individual pieces of the system.  For example, as discussed
earlier, UV-CDAT uses much of the VisTrails functionality but has a
GUI that is climate-focused.  To this end, we have had to reorganize
the code to better separate functionality from interface elements so
that they are more generally useful, enabling an \emph{elastic
  interface} that can be adapted to specific domains in order to
enhance usability.  In addition, we have worked to unentangle
functionality so that individual features can be better utilized.

\section{Outreach \& Dissemination}
\label{sec:outreach}

While workflows and provenance help scientists to be more efficient
and meticulous, such solutions work best when used at the beginning of
a project.  As work progresses, many run into issues that can be
solved with tools like VisTrails, but switching from existing
protocols is often seen as extra work that is hard to justify.  With
many research institutions now requiring full documentation of
computational analyses, there is some drive to better structure and
document experiments, but we have tried to educate potential users
about the day-to-day advantages such software provides.  To do so, we
have presented several tutorials (\eg~\cite{silva:2009:vistutorial}) about VisTrails to a
variety of audiences, demonstrated it during talks, and used it in our
own research and teaching.

In contrast to researchers who build software for their specific
domain, VisTrails was built by computer science researchers to be
useful to a variety of domains.  To demonstrate impact in computer
science, we have shown particular advantages in a variety of
areas including visualization~\cite{scheidegger:2007:analogy} and
scientific data management~\cite{koop:2010:ssdbm}.  
We have also used our work as a base for later
research.  As noted in Section~\ref{sec:collab}, other users span many
different scientific domains.  Some of our impact in domains like
climate and neuroscience has come from collaborations with domain
scientists.  However, other impact has come from users that have
independently decided to use our software for their own projects or
teaching.  While our immediate contribution is harder to quantify,
this type of impact still often requires some work from developers in
the form of documentation and e-mail support.

One of the interesting benefits of the change-based provenance in
VisTrails has been the ability to track the evolution of a solution,
and this feature has made it particularly useful in teaching.  The
software has been used to teach visualization courses, both for
presenting techniques and for students' assignments and
projects~\cite{silva:2010:eg}.  Instead of showing visualization
results as part of a lecture, VisTrails allowed lecturers to
demonstrate interactive examples that students could also download,
run, and modify.  In addition, students were required to ``show their
work'' by turning in a vistrail that showed all the steps they took in
building the visualization pipeline that solved each assigned problem.
Such information can aid instructors in understanding where students
encountered difficulties as well as common patterns used.

\section{Conclusion}

As the use of software proliferates throughout science, VisTrails
seeks to provide infrastructure for integrating varied tools while
also capturing the necessary information to ensure reproducibility.
In order to best address the needs of scientists, we have collaborated
with a number of different projects in order to both show how our work
can benefit them but also understand where new ideas are required.  As
the system has matured, we have adopted a number of standard software
engineering practices in order to better maintain and enhance it.  Our
approach has been to build software with usability in mind, addressing
key challenges as science moves to lean more on computing, and this
goal has served to provide opportunities for research in computer
science as well as end-user domains.  Finally, our outreach through
tutorials, talks, and publications has both provided opportunities
for new collaborations and contributions to communities looking to
address similar problems.

\bibliographystyle{abbrv}
\bibliography{paper}  %

\end{document}